\documentclass[letterpaper,conference]{IEEEtran}      

\usepackage{graphicx} 
\usepackage{amsmath} 
\usepackage{amssymb}  
\usepackage{mathtools}
\usepackage{algorithm}
\usepackage{algpseudocode}
\usepackage{color}
\usepackage{comment}
\usepackage{tabularx}
\usepackage{caption}
\usepackage[table]{xcolor}

\algnewcommand\algorithmicinput{\textbf{Input:}}
\algnewcommand\INPUT{\item[\algorithmicinput]}
\algnewcommand\algorithmicoutput{\textbf{Output:}}
\algnewcommand\OUTPUT{\item[\algorithmicoutput]}

\definecolor{orange}{rgb}{1,0.5,0}
\definecolor{byzantium}{rgb}{0.44, 0.16, 0.39}

\title{A framework for comprehensible multi-modal detection of cyber threats}

\author{Jan Kohout, \v Cen\v ek \v Skarda, Kyrylo Shcherbin, Martin Kopp, Jan Brabec
\\
\\Cisco Systems, Prague, Czech Republic
}

\begin{document}
\maketitle
\thispagestyle{empty}
\pagestyle{empty}

\begin{abstract}
Detection of malicious activities in corporate environments is a very complex task and much effort has been invested into research of its automation. However, vast majority of existing methods operate only in a narrow scope which limits them to capture only fragments of the evidence of malware's presence. Consequently, such approach is not aligned with the way how the cyber threats are studied and described by domain experts.
In this work, we discuss these limitations and design a detection framework which combines observed events from different sources of data. Thanks to this, it provides full insight into the attack life cycle and enables detection of threats that require this coupling of observations from different telemetries to identify the full scope of the incident. We demonstrate applicability of the framework on a case study of a real malware infection observed in a corporate network.

\end{abstract}


\section{Introduction}
There is a saying in criminology that there is no such a thing as a perfect crime since every misdemeanour leaves evidence, and cybercrime is not an exception. The clues in cyberspace are often weak, indirect, scattered around virtual locations and spread in time. Therefore, various data sources have to be analysed and adequately correlated to construct a holistic picture of a misconduct.

Imagine, for example, a security breach in a company's internal network. The incident might start by opening a phishing e-mail by one of the employees, continues with infecting the first machine and then spreads in the network using lateral movement techniques to other endpoints. The breach culminates by exfiltration of sensitive information outside the company's environment, deploying spyware tools on employees' computers or disabling company's critical infrastructure and demanding ransom. Each of these phases manifests itself by different means and is reflected in different type of telemetry data - a different \emph{modality}.

In the world of intrusion detection systems (IDS), such modalities may be logs from web proxies, file executions logs, firewall logs or e-mail data. However, the problem of multi-modal analysis is not limited to IDS only. For example, proper e-mail analysis is a multi-modal problem on its own as e-mails may contain, apart from the main textual message, hyperlinks, images, and the sender information. 

Moreover, due to the complex nature of threats, alerts generated by detection systems need to be descriptive and expressive enough such that security analysts can easily understand them and recognize the expansion of the threat in the protected environment. Again, this is possible only when information from different modalities are put together.

Capability to detect cyber threats by creating security detections spanning multiple modalities has been communicated by multiple cybersecurity vendors (e.g., \cite{paloaltonetworksXDR,CiscoXDR}) but to best of our knowledge, this is the first publication that formalises the problem and describes technical details of a framework for such multi-modal analyses. The proposed framework is not only simple and scalable but also provides comprehensible verdicts by design. The framework is demonstrated on the problem of threat detection, but it is general and can be used directly or with minor changes for other problems as well. 

This paper is organised as follows. In the next section, we present a formal definition of the problem which we address. Detailed description of the proposed framework is presented in Section~\ref{sec:framework}, followed by description of our reference implementation in Section~\ref{sec:implementation}. A case study of the framework detection capability is discussed in Section~\ref{sec:experiments}. Related work, not only from the field of cybersecurity, is presented in Section~\ref{sec:sota}.

\section{Problem Statement}
\label{sec:problem}
This section introduces the formalism needed to address the problem of security incidents detection by fusing multiple sources of data.

The primary field of interest for this study are larger networking environments in which hundreds or thousands of human users are connected using different types of devices. These may include desktop PCs and laptops used for common work, mobile devices such as phones or tablets used for communication as well as devices like printers, specialized servers (e.g., DNS, LDAP, web servers, ...), IP cameras or even very specific machines in health care or industry.

For example, a typical user in such environment uses a laptop and a smartphone to perform her daily routines. Throughout the day, a person would communicate via instant messages and e-mails, move between office floors or even buildings and access internal or external web resources. All these actions generate telemetry records of one or more types. Hence, if the person starts an e-mail client on a laptop to receive a message with a link to an external web page and later accesses it in a browser, the detection system might consume the following telemetry records: 
\begin{itemize}
\item an endpoint record that an e-mail client process was started;
\item a network record with information that the e-mail client received data from the email server;
\item a network record about the accessed external resource; 
\item an endpoint record about the browser process.
\end{itemize}

Detection of malicious activities in such setting involves to identify \emph{what} is happening (which type of threat was detected, what actions were performed), \emph{where} it is happening (which user, device or part of infrastructure is affected) and \emph{when} the detected activity occurred. In order to formally describe the detections, the following three sets are introduced - the set $Y$ of all possible types of detectable threats, the set $E$ of all \emph{entities} in the environment on which the malicious activities can be detected, and, finally, we denote $T$ a set of all possible timestamps which determines time quantization used in the detection system.

Definition of an entity can differ depending on the specific network and data available --- it can be, for example, a pair of a user and a device identifiers, or a device. The entity with examples will be discussed in more detail in Section~\ref{sec:entity_matching}.

In general, $K$ different types of input data sources can be considered which are further called \emph{modalities}. Using these modalities, the input for the detection task then consists of $K$ datasets of type

\begin{equation}
\begin{aligned}
D_k = & \{(t_i^k, d_i^k) | i \in \mathbb{N} \land i \leq n_k \} \\
      & k \in \{1, 2, \ldots, K\}, n_k = \lVert D_k \rVert,
\end{aligned}
\end{equation}

where the dataset $D_k$ originates from the $k$-th modality. The variables $t_i^k$ and $d_i^k$ represent the timestamp and the data associated with the $i$-th object in the $k$-th dataset, respectively. 

No restrictions are put in place on the structure of the data at this point as the datasets come from different modalities and thus can be completely heterogeneous. For example, if the $k$-th modality contains logs from a network sensor then the data in that dataset can be in a form of NetFlow~\cite{NetFlowsDefinition2011} records together with their timestamps of capture.

Let us denote $\mathcal{D}_k$ a set of all possible datasets originating from the $k$-th modality. The goal is to design a system that implements a function

\begin{equation}
\label{eq:system_def}
f : \mathcal{D}_1 \times \mathcal{D}_2 \times \ldots \times \mathcal{D}_K \mapsto 2^{T \times E \times Y} 
\end{equation}

In summary, the function $f$ takes $K$ datasets from the available modalities as an input and outputs a list of detections, where each detection contains information about the type of threat, affected entity and time of detection. 

 In practice, the output can also contain metadata (e.g. transferred bytes or a registry key that was changed) associated with the detection that might be useful for a security analyst working with the system's output to verify the detection and to take remediation actions.

\section{Proposed Framework}
\label{sec:framework}
Given the formal introduction of the detection problem in Section~\ref{sec:problem}, we now build a theoretical framework which enables us to describe the proposed detection system.

We see the entire system which is represented by the function  $f$ in~\eqref{eq:system_def} as a composition of functions realising individual phases of the detection process. First, detections on individual modalities are generated using modality-specific (unimodal) detectors and representations of affected entities. Second, results from the unimodal detectors are matched to assign their outputs to the same entities, we call this phase \emph{entity matching}. Finally, detections for each entity are produced based on this unified data by a multi-modal detector.

In the next subsections, we formally describe the individual phases and the composition of the final multi-modal detector.

\subsection{Unimodal detectors}
The datasets $D_k$ are passed as an input to modality-specific detectors $\delta_k$ that can be viewed as a transformation function:
$$
\delta_k: \mathcal{D}_k \mapsto 2^{T \times E_k \times \mathcal{O}},
$$
where $\mathcal{D}_k$ and $T$ are the set  of all possible input datasets and timestamps, respectively, as defined in Section~\ref{sec:problem}, and $E_k$ is the set of all entities in the environment identified using the data available in $D_k$. $\mathcal{O}$ is an \emph{event dictionary} which contains all possible detections that can be observed in the environment, given the input modalities. For simplicity, we define $\mathcal{O}$ as an universal dictionary shared across all detectors which aggregates all possible detections. 

In practice, an event is not a single element but rather a rich structure defined by its type heavily depending on the detection. E.g., for port bursts or bruteforcing attempts the event would contain quantitative information, for detection of anomalous file type the event would contain the type of the file, etc. 

Given the definitions above, the detector $\delta_k$ transforms the input dataset $D_k$ to a set of triplets: 
$$S_k = \{(t, e^k, o) | t \in T, e^k \in E_k, o \in \mathcal{O} \},$$

which contains individual detections generated by the detector $\delta_k$ using the modality $D_k$,  each associated with an affected entity and a timestamp. For better readability, we further denote the set $2^{T \times E_k \times \mathcal{O}}$ of all possible outputs $S_k$ from the detector $\delta_k$ as $\mathcal{S}_k$.

In general, the unimodal detectors can be almost arbitrarily complex and each of them can even be an aggregation of multiple detectors working on top of the given modality. However, the output  for each modality is always in a form of triplets described above thanks to which we can treat each of the unimodal detectors as a single detector, no matter what is its internal structure.

\subsection{Entity matching}

Next, results $S_k$-s from the individual modalities need to be merged with respect to the entities they refer to. Specifically, sets of detections for individual entities $e \in E$ are put together to create a complete picture of what is happening on each entity. Therefore, there needs to be a function $\gamma$ which takes outputs $S_k \in \mathcal{S}_k$ from individual unimodal detectors and outputs a set of all detections for each entity $e \in E$:
$$
\gamma: \mathcal{S}_1 \times ... \times \mathcal{S}_K \mapsto 2^{T \times E \times \mathcal{O}},
$$
The function $\gamma$ thus performs the \emph{entity matching} - i.e., deciding for two entities' representations $e^k \in E_k, e^l \in  E_l$ whether they represent the same entity $e \in E$ or not.

Again, specific implementation of the entity matching function may depend on the data available for entities in individual modalities and will be further discussed in Section~\ref{sec:entity_matching}.

\subsection{Multi-modal detector}
Given the merged outputs from unimodal detectors, the system now has enough evidence to identify suspicious actions on individual entities captured in different modalities. Therefore, the multi-modal detector can be viewed as a mapping $\tau$ from the merged results to the final threat detections: 
$$
\tau: 2^{T \times E \times \mathcal{O}} \mapsto  2^{T \times E \times Y}
$$

\begin{figure}[h]
    \centering
    \includegraphics[width=0.4\textwidth]{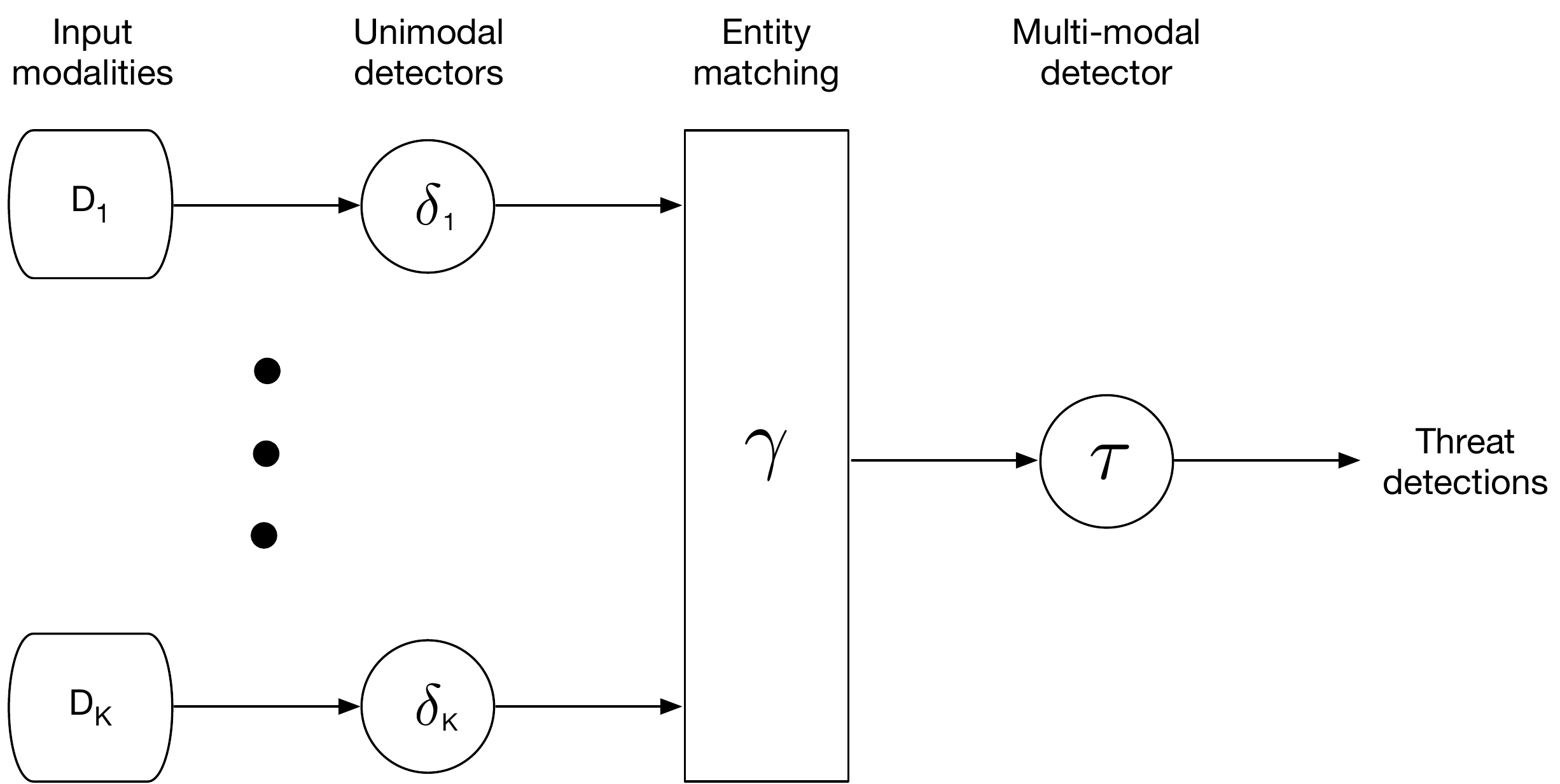}
    \caption{Composition of the proposed framework for multi-modal threat detections.} 
    \label{fig:framework_concept}
\end{figure}

Having all phases of the detection process modelled by mappings described above, we can see the multi-modal threat detector $f$ defined in Equation~\eqref{eq:system_def} as a composition of mappings from individual phases:
$$
f(D_1,...,D_K) = \tau(\gamma(\delta_1(D_1),...,\delta_K(D_K)))
$$
This concept of the entire framework is illustrated in Figure~\ref{fig:framework_concept}. The next section then describes a reference implementation built on top of the proposed framework.

\section{Implementation}
\label{sec:implementation}

\begin{figure}[h]
    \centering
    \includegraphics[width=0.4\textwidth]{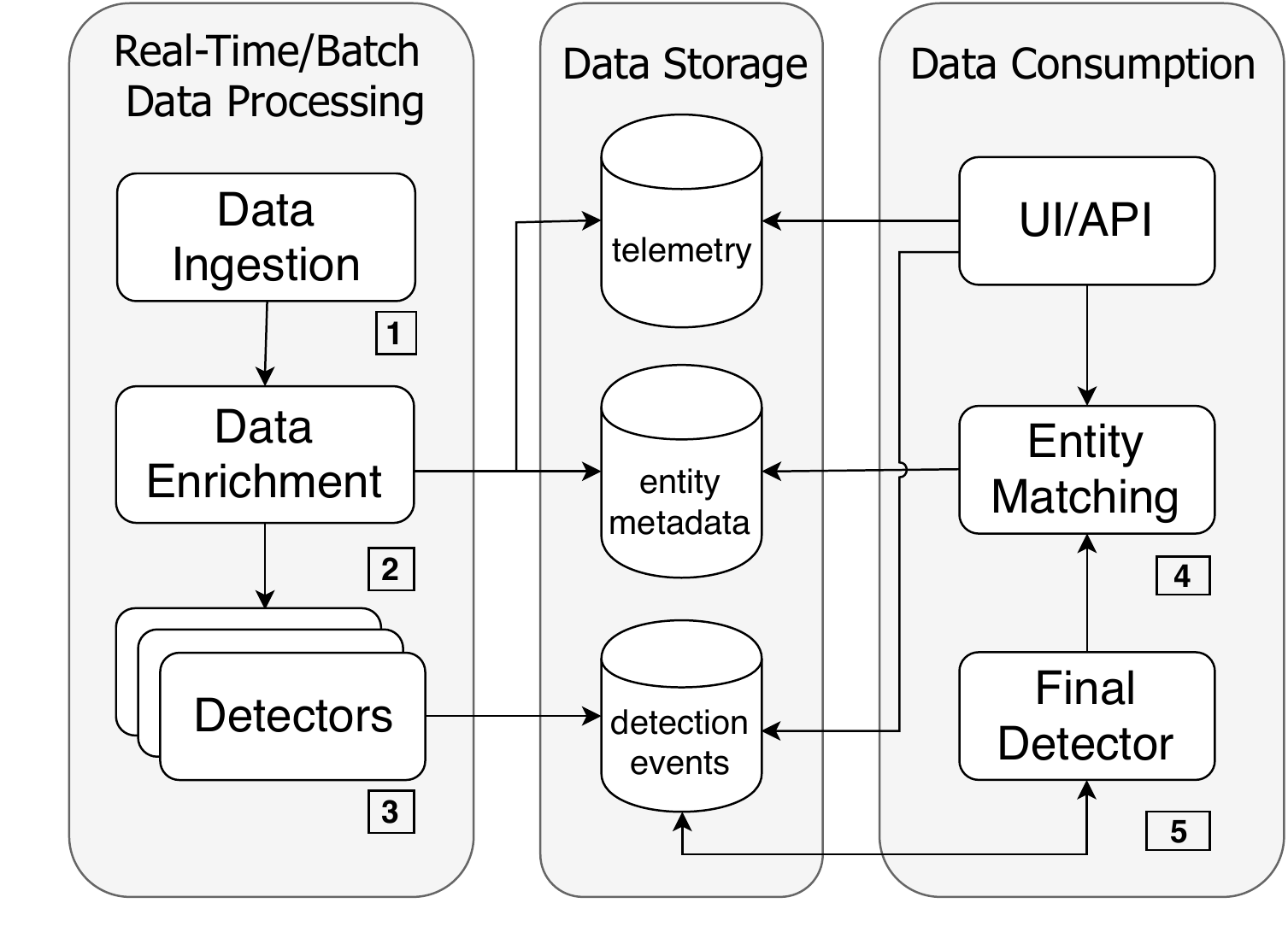}
    \caption{The reference implementation architecture.}
    \label{fig:ref-impl-architecture}
\end{figure}

This section outlines a reference implementation of a system as per the proposed framework described in Section~\ref{sec:framework}. As shown in Figure~\ref{fig:ref-impl-architecture}, the overall architecture consists of 5 components: the data ingestion and enrichment, the modality-specific detectors, the entity matching service, and the final multi-modal detector. In the following subsections, we will introduce each component in detail.

\subsection{Data ingestion and enrichment}
\label{sec:data_ingestion}

The ingest component serves as the gateway to the system. It integrates with the source data providers using both push-based and pull-based strategies. The data in various formats are validated, filtered, transformed, and enhanced with contextual or global intelligence information. Finally, the data is catalogued and persisted to the data lake for further reference by the downstream components and human analysts. 

The reference system was implemented with two modalities: network telemetry and endpoint telemetry. The network data used are additionally of two types: web proxy logs and NetFlows, which include fields such as IP addresses for the client and the contacted server, number of transferred bytes and packets, timestamp, duration of the communication, and URL or  SNI for the HTTP and HTTPS connections, respectively. 

The endpoint telemetry is provided by lightweight connectors installed on the client devices (laptops/PCs, tablets and smartphones). These logs are substantially more heterogeneous than the network logs. Different schema with some common fields (such as timestamp and endpoint ID) is used for logs of file creation, execution of the binary, download of the binary from the web, file transfer, etc.

\subsection{Unimodal detectors}
\label{sec:impl_unimodal_detectors}

Unimodal detectors can implement various detection methods developed for individual modalities. Besides traditional rule-based systems which generate events based on matched hard-coded patterns, they might include anomaly detection algorithms or supervised classifiers. The large variety of anomaly detection methods applicable in the security domain can employ simple statistical detectors such as~\cite{eskin2002geometric,kim2004packetscore} or probabilistic  models~\cite{goldstein2012histogram, kopp2018community}. For more complete survey on anomaly detection algorithms, see, for example,~\cite{outliers, adSurvey}.

For the pretrained variants of unimodal detectors, one can consider random forest classifiers which proved to be very effective in  many security applications~(\cite{brabecForest, hansenForests}), similarly as artificial neural networks~(\cite{kalash2018malware, khan2019improved}). However, even much simpler yet computationally effective methods are relevant here because even weaker detections can contribute to the final detections produced by the multi-modal detector.

At the time of writing this paper, the reference system included approximately 300 unimodal detectors falling into four categories:

\begin{itemize}
\item \emph{signature-based} -- producing detection events by matching behavioral signatures handcrafted by a domain expert;
\item \emph{classifier-based} -- generating events by supervised classifiers trained on historical data;
\item \emph{anomaly-based} --  leveraging  statistical, volumetric, proximity, targeted, and domain specific anomaly detectors;
\item \emph{contextual events} -- capturing various network or endpoint behaviors to provide additional context, e.g., file download events, direct access on raw IP or software updates.
\end{itemize}

Each detector transforms the data source according to its needs and emits an event if the triggering criteria are met. The detector persists each event to the events store, where it can be later evaluated by the multi-modal detector to produce a final security detection.

\subsection{Entity matching}
\label{sec:entity_matching}

As mentioned earlier, the same entity can be observed simultaneously in multiple modalities, represented by different identifiers. It is the Entity Matching (EM) component's prime responsibility to link the data sources together. The component examines the metadata about entities and the telemetry feeds to construct a unified view of the entities in the observed environment's local and global contexts.

The unimodal detectors process the telemetry records independently and potentially produce the events. At the same time, the EM component using the directory, inventory, and other metadata services would attempt to establish the relationships among the observed records. Generally, it would associate the person to their laptop, the laptop to the email and browser processes, their corresponding assigned network addresses, thus linking the telemetry and the generated events to a single entity. This entity is referred to as a \emph{cross-modal entity}.

In the reference implementation, the EM constructs the mapping between endpoint IDs and IP addresses. The EM algorithm prefers precision over the completeness of IP space coverage, meaning that IP addresses used by multiple endpoints concurrently (e.g., endpoints located in different subnets with the same IP ranges) is discarded.

This mapping is used to associate detections from network modalities with given IP address and timestamp to the cross-modal entity. In this reference implementation, the endpoint-modality entity is equal to the cross-modal entity --- the unique endpoint ID.

\subsection{Multi-modal detector}
\label{sec:multi_modal_detector}

The multi-modal detector is a component which operates on observations from bounded time window and multiple modalities that were already associated with the cross-modal entity. The detector can be as simple as a single rule mapping specific observation directly on a threat detection. This approach is sufficient if the observation is already significant enough on its own. However, some threats can be identified with high probability only by correlation of multiple observations from multiple modalities. In this case, the threat detector can be defined by a complex rule or set of rules. The rule itself may be explicit or encoded by e.g. a neural network.

Generally, we are solving a classification task on top of chronological and parametric itemsets \cite{dauxais2017discriminant}. These itemsets can be labelled by threats based on some threat intelligence which is mostly represented by a set of malicious binaries or network domains, phishing e-mails, etc. The labelling can be associated with signature-based observations as well as extrapolated to the cross-modal entity. In such scenario, the task leads to a multi-label classification task. 

When proposing the model the main concerns are:
\begin{itemize}
    \item \emph{extractability}, the existence of efficient algorithm,
    \item \emph{interpretability}, the understandability of extracted behaviours by domain experts,
    \item \emph{expressiveness}, the discriminative power of the model,
    \item and \emph{generality}, the ability of the algorithm to be retrained on new observations without significant changes.
\end{itemize}
These properties can not be optimised all together due to the contradictory nature of expressiveness and interpretability of the model.

Nowadays, interpretability is often required by threat researchers and treat response teams. Even precise detection without a proper "story" behind might get overseen by threat response teams due to lack of reliability and the amount of work needed to confirm or disprove the detection. Due to this, models built on top of the statistical based rule mining are used in our current reference implementation.

The rule mining algorithm mines rules from itemsets -- items grouped by a key. In our case, the items are detections from unimodal detectors and the key is the cross-modal entity. If all itemsets were processed together, only the noise from unimodal detectors would be mined as it is hopefully the only common behaviour across all entities.

Due to this, the data are split into subgroups based on labels assigned to entities by threat intelligence. Therefore, the expectation is that each group contains similar behaviours which are rarely seen outside this group. 

In the first step of the algorithm, frequent itemsets are mined in each group by the FP-Growth algorithm \cite{wang2002top}. This step returns frequent rules in each group but it doesn't ensure that they are not frequent also in other groups. In the next step we filter out those which are. Finally, association rules correlated with one group only are obtained.

Next, the multi-modal detector interpreting association rules is defined. It has to deal with the possibility of matching an entity with rules from multiple groups. In our case, the group with more matching rules is chosen as the label for each entity. In case of a tie, the group corresponding to the more severe threat is chosen.

\section{Case study} \label{sec:experiments}
Here we present preliminary detection statistics from the reference system with emphasis on a specific example of combined detection which illustrates the need to combine events from different modalities to successfully identify the malware.

The system was verified on 30 days of telemetry data from 33 real corporate networks containing 758.043 cross-modal entities on which security analysts identified 73 different threats.

Table \ref{tab:overall_efficacy} shows comparison of precision and recall between rules mined over events from individual modalities and combined modalities. We emphasise that the absolute values of precision and recall are not of the main interest here - it is the capability of the proposed system to detect new malware families based on combining events from different modalities (those malware families are highlighted in green).

\begin{table}[h!]
    \centering
    \begin{tabular}{l|r|r|r|r|r|r}
        malware & \multicolumn{2}{c|}{endpoint} & \multicolumn{2}{c|}{network} & \multicolumn{2}{c}{combined} \\ 
        & pre & rec & pre & rec & pre & rec \\ \hline
        Ad injector & 0.25 & 0.25 & 0.79 & 0.44 & 0.79 & 0.44 \\
        PUA Genieo & nan & 0.00 & 0.50 & 0.12 & 0.30 & 0.16 \\
        PUA Trojan.Patchbrowse & nan & 0.00 & 1.00 & 0.19 & 1.00 & 0.19 \\
        PUA Pirrit & nan & 0.00 & 0.88 & 0.28 & 0.88 & 0.28 \\
        PUA ArcadeYum & nan & 0.00 & 0.43 & 0.44 & 0.43 & 0.47 \\
        PUA Conduit & nan & 0.00 & 0.97 & 0.26 & 0.97 & 0.26 \\
        PUA Crossrider & nan & 0.00 & 1.00 & 0.75 & 1.00 & 0.75 \\
        Fake Search Engine & nan & 0.00 & 0.82 & 0.33 & 0.82 & 0.33 \\
        Downloader & nan & 0.00 & 1.00 & 0.50 & 1.00 & 0.50 \\
        Malware distribution & nan & 0.00 & 0.38 & 0.23 & 0.37 & 0.23 \\ 
        Worm & 0.50 & 0.16 & nan & 0.00 & 0.50 & 0.16 \\
        \rowcolor{green} 
        Dropper - OSX/Shlayer & nan & 0.00 & nan & 0.00 & 0.14 & 0.05 \\
        \rowcolor{green}
        Trojan & nan & 0.00 & nan & 0.00 & 0.83 & 0.20 \\
    \end{tabular}
    \caption{Per-entity detection statistics of the reference system for different malware families with at least one entity detected as TP or FP. }
    \label{tab:overall_efficacy}
\end{table}

We specifically illustrate the cross-domain detection capability on the OSX/Shlayer malware, whose behavioral pattern was discovered by the multi-modal detector (while it was not detectable whithin a single modality - see Table~\ref{tab:overall_efficacy}), which learned the following combination of events for its detection (listed together with the source modality):

{\small
\begin{enumerate}

\item Contacting a very long URL for the first visit of a hostname (\emph{network telemetry})
\item Unusual file hash for the given environment (\emph{endpoint telemetry})
\item Unusual file type, 7Zip files downloaded from a specific URL in this particular case (\emph{endpoint telemetry})
\item Communication using a user agent rarely used for the given site (\emph{network telemetry})
\item Connection check (\emph{network telemetry})
\item Reoccurring communication with a rarely visited site (\emph{network telemetry})
\item Communication using an unusual user agent for the given user (\emph{network telemetry})
\item Usage of device fingerprinting scripts and tools (\emph{endpoint telemetry})
\item Usage of unusual remote port (\emph{endpoint telemetry})

\end{enumerate}
}

The OSX/Shlayer malware's behavioral pattern is infecting computers by a fake flash plugin update, which is downloaded as a 7zip file (this involves contacting a very long URL for the first time and unusual hash and file type). When this update is installed, the real flash plugin is downloaded from the real update page using a \texttt{curl} command (unusual user agent for a well-known site).

When the OSX/Shlayer is on the infected device, it checks if that device is connected to the internet (connection check) and starts gathering information about the infected device (usage of fingerprinting scripts and tools). After that, it establishes a C\&C channel (reoccurring communication to a rarely visited site) and starts exfiltrating sensitive information about the infected device either via a specific URL pattern or in the user agent field (communication using an unusual user agent for the given user). 

The next patterns may differ because the OSX/Shlayer is the specimen of the Dropper malware class, therefore it has the capability to install additional modules for various purposes, e.g.: malicious advertising, click fraud or crypto mining, see~\cite{shlayer} for further reference.

\section{Related Work}
\label{sec:sota}

Outside of cybersecurity, multi-modal machine learning is an active topic with long history of research in fields such as robotics~\cite{noda2014multimodal,park2018multimodal,caltagirone2019lidar}, multimedia~\cite{yuhas1989integration,wu2006multi} and others~\cite{deo2018multi,wang2018eann, li2021multi}. Ref.~\cite{baltruvsaitis2018multimodal} provides an extensive taxonomy of challenges in multi-modal ML and approaches to design of classifiers in the more studied domains. 

Multi-modal learning in cybersecurity does not have such a long history. This may be because the modalities do not map into basic human senses as in the previous domains. There have been a handful of publications concerned with multi-modal learning in cybersecurity.

Several systems for multi-modal detection of malware in the Android ecosystem have been proposed~\cite{singh2021multi,millar2020dandroid, narayanan2018multi, appice2020clustering, millar2021multi}. In these, the modalities (permissions, opcodes, API calls, \ldots) are extracted from APK files and a classification system is trained on a dataset of labelled APK files.

Similar methods have been presented in classification of binaries~\cite{bai2016improving, guo2010malware}. Here again, all of the modalities can be extracted from a single executable file.

The modalities we consider in this work are high-level and originate from distinct data-sources (endpoint, network, email) which are decoupled into fully separate systems. This makes just the fusion of the modalities together an interesting problem as we described in Section~\ref{sec:entity_matching}. We are not aware of another publication addressing this problem.

The detection system we propose is composed of modality-specific detectors which are further combined by the multi-modal detector. In this work, we are not concerned with design of individual modality-specific detectors. For the final detector, we have decided to use rule mining approach due to comprehensibility of its output. However, almost any classification algorithm can be considered.

There exist many approaches to rule mining such as frequent item set mining e.g.~\cite{borgelt2003efficient,uno2004lcm,han2000mining,wang2002top}, logical item set mining~\cite{kumar2012logical}, sequential rule mining~\cite{dalmas2017twincle}, discriminant chronicles mining mining~\cite{dauxais2017discriminant}, or currently emerging deep rule mining~\cite{beck2021empirical}. According to the comparison of rule mining approaches~\cite{kopp2018comparing} we have selected a frequent item set mining, specifically an FP-growth algorithm~\cite{wang2002top} with slight modifications, as discussed in Section~\ref{sec:multi_modal_detector}. 

\section{Conclusion}
\label{sec:conclusion}
This work emphasizes the need of combining multiple sources of data in automated detection of security incidents in order to provide complete and descriptive detections, referred as multi-modal detections. The problem of multi-modal detection was formally defined and a theoretical framework was introduced. A reference implementation of the framework was created and used to demonstrate the detection capabilities of the multi-modal approach.

We see this contribution as a starting point and a baseline solution for future development of multi-modal detection capabilities, which will play key role in advanced detections of cyber threats with the help of AI methods.

\bibliographystyle{IEEEtran}
\bibliography{xda}	

\end{document}